\documentclass[aps,prl,twocolumn,nofootinbib,groupedaddress]{revtex4-2}

\usepackage[utf8]{inputenc}
\usepackage[T1]{fontenc}
\usepackage[english]{babel}
\usepackage{graphicx}
\graphicspath{{numerics/}{./}}
\usepackage{amsmath,amssymb,amsbsy,amstext,amsfonts}
\usepackage{bm}
\usepackage{xcolor}
\usepackage{hyperref}
\hypersetup{colorlinks=true,linkcolor=purple,filecolor=purple,urlcolor=purple,citecolor=purple}

\renewcommand{\d}{\mathrm{d}}

\renewcommand{\Im}{\mathrm{Im}}
\newcommand{\rhohat}{\hat{\rho}}        %
\newcommand{\mhat}{\hat{m}}             %
\newcommand{\nueff}{\nu_{\rm eff}}      %
\newcommand{\mueff}{\mu_{\rm eff}}      %
\newcommand{\pic}{\pi_c}                %

\newcommand{\Ga}{\Gamma}
\newcommand{\FF}[2]{{}_{#1}F_{#2}}
\newcommand{\lamstar}{\lambda_{\star}}

\newcommand{\SM}{the Supplemental Material}

\begin{document}

\title{Analytical Cosmological Collider at Strong Mixing in Laplace Space}

\author{Nathan Belrhali}
\author{Arthur Poisson}
\author{S\'ebastien Renaux-Petel}
\affiliation{Institut d'Astrophysique de Paris, CNRS, Sorbonne Universit\'e,
98 bis bd Arago, 75014 Paris, France}

\begin{abstract}
Primordial non-Gaussianities are at their largest when the curvature
perturbation is coupled to a massive field by a mixing too strong
to be treated perturbatively. We show that this regime, naturally reached in
multifield inflation, is solved analytically by the Laplace-space representation we recently
introduced, in which every mode becomes a superposition of plane waves weighted
by a single kernel.
The bispectrum then follows as a rapidly convergent series of elementary functions, in perfect agreement with independent numerical results.
We find that the cosmological collider signal at strong mixing is larger than widely believed, with only half of the usual Boltzmann suppression. Its amplitude further grows with the number of time derivatives of the curvature perturbation in the interaction, singling out the most promising observational targets.
\end{abstract}

\maketitle

\paragraph*{\bf Introduction.}
The correlators of the primordial curvature perturbation carry the imprint of
every field present during inflation. A field with a mass of the order of the
Hubble scale leaves a distinctive signature: its spontaneous production
makes the squeezed limit of the bispectrum oscillate in the
logarithm of the momentum ratio, with a frequency that measures the mass. Detecting
this oscillation, and through it the particle content of inflation, is the
goal of the cosmological collider program~\cite{Chen:2009we,Chen:2009zp,Baumann:2011nk,Noumi:2012vr,Arkani-Hamed:2015bza,Lee:2016vti}. Once a theoretical proposal, it is becoming an observational target:
dedicated searches in cosmological data have
begun~\cite{Sohn:2024xzd,Cabass:2024wob,Philcox:2025bbo,Suman:2025vuf,Kumar:2026ogn,Philcox:2026tjj}.

The size of the signal rests on the coupling between the massive field and
the curvature perturbation. When several fields are present, the inflationary
trajectory generically turns, and the turn generates a linear mixing between
the two. Nothing requires it to be weak. At strong
mixing, the signal is at its largest~\cite{Werth:2023pfl,Pinol:2023oux,Jazayeri:2023xcj}. But no expansion in
powers of the mixing converges, and the regime long seemed out of analytic
reach. It has recently triggered intense
activity~\cite{Kumar:2026ogn,Huenupi:2026abj,Huenupi:2026aqc,Pinol:2026xnl,Wang:2026lff,Philcox:2026rpn,Philcox:2026tjj}, from exact solutions of the linear theory to numerical evaluations of
the bispectra and first confrontations with data.

This Letter solves the strongly mixed regime analytically, from the linear
theory to the bispectrum. Our route is independent of these works: it
continues the Laplace-space approach we recently developed~\cite{Belrhali:2026ygh,Belrhali:2026jqe}, in which every mode function is a superposition of flat-space plane waves, with
the field content and dynamics carried entirely by the kernel of that superposition. The same representation solves the strongly mixed system.
The linear theory comes out exactly, with the late-time curvature and the
power spectrum in closed form. The bispectrum then follows as a rapidly
convergent series involving only elementary functions, almost instantaneous to evaluate. It agrees perfectly with the independent numerical results of~\cite{Philcox:2026tjj}, obtained with CosmoFlow~\cite{Werth:2024aui}.

These analytical results show that the amplitude of the cosmological collider signal is larger at strong mixing than widely believed, with only half of the usual Boltzmann suppression. They also reveal a hierarchy among the interactions: each time derivative of the curvature perturbation enhances the oscillations by one power of their frequency. For the operators with the most time derivatives, the collider signal therefore stands out most clearly above the smooth background.

\paragraph*{\bf Setup.}
We work in quasi-de Sitter space and at unit sound speed, in the decoupling limit of the
effective field theory of inflationary fluctuations~\cite{Creminelli:2006xe,Cheung:2007st},
extended by a massive scalar $\sigma$~\cite{Noumi:2012vr,Pinol:2023oux,Pinol:2024arz}. The curvature sector is described by the canonically normalized Goldstone
boson $\pic$, related to the curvature perturbation by $\zeta=-H\pic/f_\pi^{2}$, with $f_\pi^{4}=2M_{\rm Pl}^{2}|\dot H|$ the symmetry breaking scale.
It couples to $\sigma$, of mass $m$, through the linear mixing
$\rho\,\dot{\pi}_c\,\sigma$, with $\rho$ constant. At leading order in derivatives, the Lagrangian up to cubic order
reads~\cite{Pinol:2023oux}
\begin{equation}
\label{eq:lagrangian}
\begin{aligned}
&\mathcal L/a^{3}=\frac12\dot{\pi}_c^{2}-\frac{(\partial_i\pic)^{2}}{2a^{2}}
+\frac12\dot\sigma^{2}-\frac{(\partial_i\sigma)^{2}}{2a^{2}}-\frac12 m^{2}\sigma^{2}\\
&+\rho\,\dot{\pi}_c\,\sigma-g\,\sigma^{3}-\frac{\alpha}{2}\,\dot{\pi}_c\,\sigma^{2}-\frac{\dot{\pi}_c^{2}\sigma}{2\Lambda_2}-\lambda_2\,\dot{\pi}_c^{3}-\frac{(\partial_i\pic)^{2}\sigma}{2\Lambda_1 a^{2}}\,.
\end{aligned}
\end{equation}
This Letter considers arbitrary values of the mass and the mixing, and in particular the regime of \emph{strong mixing},
$\rhohat\equiv\rho/H\gtrsim\mhat\equiv m/H$, where no expansion in powers of $\rhohat$ converges.
At quadratic order, in the variable $z=-k/(a H)<0$ and with primes denoting
$\d/\d z$, each Fourier mode obeys
\begin{align}
\label{eq:system}
\pic''-\frac{2}{z}\,\pic'+\pic&=-\frac{\rhohat}{z}\,\sigma'
  +\frac{3\rhohat}{z^{2}}\,\sigma\,,\notag\\
\sigma''-\frac{2}{z}\,\sigma'+\Big[1+\frac{\mhat^{2}}{z^{2}}\Big]\sigma
  &=\frac{\rhohat}{z}\,\pic'\,.
\end{align}

\paragraph*{\bf Linear theory from Laplace space.}

How to quantize the system is well known, most simply in the two combinations
$\pic-i\,a\,\sigma$, $a=\pm1$~\cite{Assassi:2013gxa,An:2017hlx}, which diagonalize the sub-horizon evolution: deep
inside the horizon, each evolves as
\begin{equation}
\label{eq:earlytime}
\pic-i\,a\,\sigma\ \propto\ z\,e^{-iz}\,(-z)^{-a\,i\rhohat/2}\,,
\end{equation}
the mixing entering through the logarithmic phase. We call these two degrees
of freedom the \emph{channels} of the mixed system, and expand both fields on the
same pair of quanta,
\begin{equation}
\label{eq:quanta}
X_{\bm k}=\sum_{a=\pm1}\Big[\,X_a(z)\,\hat b_{a,\bm k}
+X^{*}_a(z)\,\hat b^{\dagger}_{a,-\bm k}\Big]\,,
\qquad X=\pic,\sigma\,,
\end{equation}
with $[\hat b_{a,\bm k},\hat b^{\dagger}_{b,\bm k'}]
=(2\pi)^{3}\delta_{ab}\,\delta^{(3)}(\bm k-\bm k')$. The mode functions
$(\pi_{c,a},\sigma_a)$ form the channel-$a$ solution of \eqref{eq:system}, with the Bunch--Davies initial
conditions $\pi_{c,a}=-i\,a\,\sigma_a\,[1+\mathcal O(1/z)]$ ensuring the early-time behavior \eqref{eq:earlytime}. Two-point correlation functions are then given by a sum over
channels, $\langle X_{\bm k}(z_1)Y_{\bm k'}(z_2)\rangle
=(2\pi)^{3}\delta^{(3)}(\bm k+\bm k')\sum_{a}X_a(z_1)Y^{*}_a(z_2)$.

The channels are solved exactly by the Laplace representation developed in our
previous work~\cite{Belrhali:2026ygh,Belrhali:2026jqe}: a Bunch--Davies mode function can be written as a continuous
superposition of flat-space plane waves,
\begin{equation}
\label{eq:sigmaray}
\sigma_a(z)=\sqrt{w_a}\;z^{2}\!\int_1^{\infty}\!\d\lambda\;
B_a(\lambda)\,e^{-i\lambda z}\,,
\end{equation}
in which the plane wave carries the entire time dependence while everything
specific to the theory sits in the kernel $B_a(\lambda)$, the discontinuity of
the Laplace transform of the mode function across its branch cut. The endpoint
$\lambda=1$ of the Laplace-frequency domain is the dual image of the early-time
Bunch--Davies behavior. The factor $z^{2}$ is a convenient choice
that keeps the kernel bounded there, and $w_a$ is a normalization
constant fixed below. The kernel is determined by a dual equation of motion:
under \eqref{eq:sigmaray}, an equation with coefficients polynomial
in $z$ becomes a differential equation in $\lambda$, of order set by the highest
power of $z$. The key simplification is that \emph{the transform itself is never
needed, only its discontinuity}. Eliminating $\pic$ from \eqref{eq:system} leaves
a single fourth-order equation in time for $\sigma$, with coefficients of degree
four in $z$ once put in polynomial form. Its dual is therefore a fourth-order
operator in $\lambda$, which turns out to be an exact second derivative,
$\d^2/\d\lambda^2\circ\mathcal L_2$, of the second-order operator
\begin{multline}
\label{eq:L2}
\mathcal L_2[B]=(\lambda^{2}-1)^{2}B''+2\lambda(\lambda^{2}-1)B'\\
+\big[(\mhat^{2}+\rhohat^{2}-2)\lambda^{2}-(\mhat^{2}-2)\big]B\,.
\end{multline}
The dual equation therefore implies that $\mathcal L_2$ applied to the
transform is a first-degree polynomial in $\lambda$. Since polynomials have
no discontinuity, the kernel itself obeys the simple second-order equation
$\mathcal L_2[B_a]=0$. This is how the Laplace representation
solves the strongly mixed system: a two-field problem, of fourth order once
decoupled, reduces to a single second-order equation. The solution that reproduces the
early-time phase $(-z)^{-a\,i\rhohat/2}$ of channel $a$ reads
\begin{equation}
\label{eq:kernel}
\begin{aligned}
&B_a(\lambda)=\Ga(1-2s)\,P^{\,2s}_{\nueff-\frac12}(\lambda)\,,\\
& \nueff^{2}=\frac94-(\mhat^{2}+\rhohat^{2})\,,\quad   s=-\frac{a\,i\rhohat}{2}\,.
\end{aligned}
\end{equation}
Remarkably, the Laplace-space representation of the massive mode function in the full coupled system takes the same simple form as the one for a free field~\cite{Belrhali:2026ygh,Belrhali:2026jqe}.
 The mixing only promotes the kernel to an associated Legendre function of degree $2 s$, and shifts the order, now governed by the effective mass $m_{\rm eff}^2=m^{2}+\rho^{2}$, well understood to characterize the system~\cite{Castillo:2013sfa,An:2017hlx,Iyer:2017qzw,Werth:2023pfl,Pinol:2023oux}. Since $s$ is purely
imaginary and $\nueff^{2}$ is real, one has $B^{*}_a=B_{-a}$ on $(1,\infty)$. The constant $w_a$ is in turn fixed by the
canonical commutation relations, which give
\begin{equation}
\label{eq:weights-w}
w_a=\frac{H^{2}}{2k^{3}}\,\frac{\sinh(\pi\rhohat/2)}{\pi\rhohat}\;
e^{+a\pi\rhohat/2}\,,
\end{equation}
the analogue, for each channel, of the familiar normalization
$H/\sqrt{2k^{3}}$ of a free field, the two coinciding as
$\rhohat\to0$. Finally, the equations of motion determine $\pic$ algebraically from $\sigma$, and one finds
\begin{multline}
\label{eq:picray}
\pi_{c,a}(z)=-\frac{\sqrt{w_a}}{\rhohat}\!\int_1^{\infty}\!\d\lambda\;
B_a(\lambda)\,e^{-i\lambda z}\Big[i\lambda(\lambda^{2}-1)z^{3}\\
+(1-3\lambda^{2})z^{2}
-\big(\mhat^{2}+\rhohat^{2}-2\big)(1+i\lambda z)\Big]\,.
\end{multline}
The single Laplace-space kernel \eqref{eq:kernel} is thus enough to generate the four mode functions of the
coupled system.

To read off the late-time values of the curvature mode functions $\pi_{c,a}(0)$, it is useful to introduce the \emph{moment function} of the kernel:
\begin{equation}
\label{eq:momentfunction}
M_a(u)=\int_1^{\infty}\!\d\lambda\;\lambda^{-u}B_a(\lambda)\,,
\end{equation}
closely related to its Mellin transform, defined by
the integral where it converges and by analytic continuation elsewhere. A closed form is known for arbitrary $u$, and the only ingredients needed in what follows are the \emph{moments} $M_a(-n)$, for integers $n$, which reduce to Gamma functions alone. Taking the $z\to0$ limit of \eqref{eq:picray} gives $\pi_{c,a}(0)=\sqrt{w_a}\,(\tfrac14-\nueff^{2})M_a(0)/\rhohat$, from which the power
spectrum follows: $\Delta_\zeta^{2}=R\,\Delta_{\zeta,0}^{2}$ with the uncoupled value $\Delta_{\zeta,0}^{2}=(4 \pi^2)^{-1}(H/f_\pi)^4$ and the amplification factor $R$ derived in \SM, in agreement with~\cite{Huenupi:2026abj,Pinol:2026xnl}. Since
$B^{*}_a=B_{-a}$, the ratio of the
late-time values is fixed by the normalizations alone,
\begin{equation}
\label{eq:asymmetry}
\frac{\lvert\pi_{c,+}(0)\rvert}{\lvert\pi_{c,-}(0)\rvert}
=\sqrt{\frac{w_{+}}{w_{-}}}=e^{+\pi\rhohat/2}\,,
\end{equation}
independently of the mass. The two $\pm$ channels are equally excited in the vacuum, but the mixing exponentially amplifies the $+$ quanta relative to the $-$ ones, much like the axial coupling amplifies one helicity over the other in axion-gauge field inflation~\cite{Anber:2009ua}.
The same is not true of $\sigma$: for $m>3H/2$, for instance, the envelope
hierarchy is reversed, with the $-$ channel dominating by $e^{\pi\rhohat/2}$.

At large $\lambda$
the kernel decays as the sum of two powers,
$B_a\simeq D^{(a)}_{+}\lambda^{-\frac12+\nueff}
+D^{(a)}_{-}\lambda^{-\frac12-\nueff}$, with coefficients again given by ratios
of Gamma functions,
\begin{equation}
\label{eq:tails}
D^{(a)}_{\pm}=\frac{\Ga(1-2s)\,\Ga(\pm2\nueff)\;2^{\frac12\mp\nueff}}
{\Ga\!\left(\tfrac12\pm\nueff\right)\Ga\!\left(\tfrac12\pm\nueff-2s\right)}\,.
\end{equation}
Under the representations \eqref{eq:sigmaray} and \eqref{eq:picray}, these two \emph{tails} are the dual images
of the two late-time behaviors $(-z)^{3/2\pm\nueff}$, dominant for $\sigma$ and subleading for $\pic$.
They encode the production of particle pairs in the system, and at strong mixing they play a role
conceptually analogous to that of Bogoliubov coefficients for a heavy free field.

\paragraph*{\bf Bispectrum from Laplace space.}

Consider first the $\sigma^{3}$
interaction. At tree level, the in-in formula gives
\begin{equation}
\label{eq:inin}
\big\langle\pic^{3}\big\rangle'
=12\,g\;\mathrm{Im}\int_{-\infty}^{0}\!\frac{\d\tau}{(H\tau)^{4}}\,
\prod_{i=1}^{3}\Big[\sum_{a_i}\pi_{c,a_i}(0)\,\sigma^{*}_{a_i}(k_i\tau)\Big]\,,
\end{equation}
with $\tau=-1/(aH)$ the conformal time, so that $z=k\tau$, and
where each of the three bracketed factors, a \emph{leg} of the correlator, is
the two-point function that connects the late-time curvature to the field at
the vertex. Rotating the time contour onto the Euclidean axis turns each
conjugated mode function into the Laplace
transform of the kernel $B_{-a_i}$. The remaining time integral is
elementary,
$\int_0^\infty\d t\,t^{2}e^{-Kt}=2/K^{3}$, and one obtains
\begin{equation}
\label{eq:assembly}
\big\langle\pic^{3}\big\rangle'
=\frac{24\,g}{H^{4}}(k_1k_2k_3)^{2}\,
\mathrm{Im}\Big[i\sum_{\vec a}
\prod_{i=1}^{3}\sqrt{w_{a_i}}\,\pi_{c,a_i}(0)\,
\mathcal J_{-\vec a}\Big],
\end{equation}
with $\vec a=(a_1,a_2,a_3)$ labeling the channels of the legs, and
\begin{equation}
\label{eq:master}
\mathcal J_{\vec a}=\iiint_1^\infty \prod_{i=1}^{3}\d\lambda_i\,
B_{a_i}(\lambda_i)\;\frac{1}{K^{3}}\,,\qquad K=\sum_i k_i\lambda_i\,,
\end{equation}
the \emph{vertex integral}: the three kernels integrated against the deformed
total energy denominator, each momentum rescaled by its Laplace frequency. The computation is structurally the same for all interactions: $1/K^{3}$ is replaced by a short list of monomials in the
$\lambda_i$ divided by other powers of the same denominator, and the mixed
interactions distribute the three momenta differently among the legs.

\begin{figure*}[t]
\centering
\includegraphics[width=\textwidth]{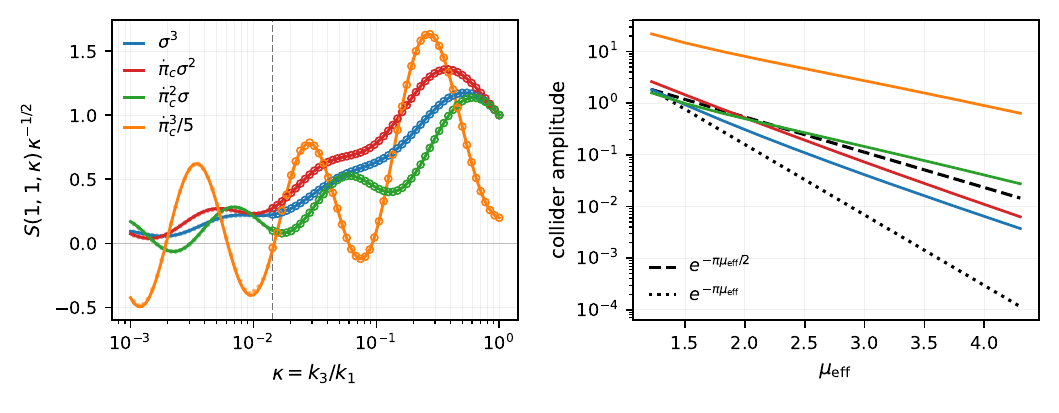}
\caption{\emph{Left}: shapes of the strongly mixed bispectrum along the isosceles family,
rescaled as $S(1,1,\kappa)\,\kappa^{-1/2}$, for
the first four cubic interactions in \eqref{eq:lagrangian}, each normalized at the
equilateral configuration, at $\mhat^{2}=1.25$ and $\mueff \equiv -i \nueff= 3$. The
$\dot\pi_c^{3}$ curve is divided by $5$ to fit the panel. Open circles are the results
of~\cite{Philcox:2026tjj},
obtained with CosmoFlow~\cite{Werth:2024aui}: the most squeezed configuration computed lies at the dashed line, further squeezing
being numerically prohibitive and handled by a fitted form, shown as small dots. \emph{Right}: amplitude of the
cosmological collider oscillations of each interaction as a function of $\mueff$ at fixed $\mhat^{2}=1.25$, normalized by its own shape at the equilateral configuration, together with the Boltzmann and half-Boltzmann scalings. All interactions follow the half-Boltzmann envelope, with power-law prefactors growing with the number of $\dot\pi_c$ legs.}
\label{fig:shapes}
\end{figure*}

To evaluate \eqref{eq:master} and its analogues, we order $k_3\le k_2\le k_1$. The softest leg and
the two hard legs are treated differently: the former is expanded in a
convergent series, the latter two integrated together. The soft Laplace frequency $\lambda_3$ enters only
through the \emph{soft-leg integral}
\begin{equation}
\label{eq:softblock}
T_{\beta}(\lamstar)=\int_1^{\infty}\!\d\lambda\;
\frac{B_a(\lambda)}{(\lambda+\lamstar)^{\beta}}\,,
\qquad \lamstar=\frac{k_1\lambda_1+k_2\lambda_2}{k_3}\,,
\end{equation}
with $\beta=3$ for \eqref{eq:master}. The variable $\lamstar \geq 2$ is itself a
Laplace frequency, the one at which the hard pair is seen by the soft leg:
the total energy is $K=k_3(\lambda_3+\lamstar)$, so the pair enters exactly
as would a single plane wave of momentum $k_3$ and frequency $\lamstar$.
What organizes the expansion is the
analytic structure of the kernel, through its moment function
\eqref{eq:momentfunction}. Intuitively, frequencies below $\lamstar$ probe the low-energy content and generate integer powers of
$1/\lamstar$, while frequencies above
$\lamstar$ probe the high-frequency behavior of the kernel, that is, the pair-creation content
of the leg, carried by its two tails
\eqref{eq:tails} and their subleading corrections. This gives the structure, derived in \SM,
\begin{equation}
\label{eq:towers}
T_{\beta}(\lamstar)=\sum_{n\ge0}t_{n}\,\lamstar^{-\beta-n}
+\sum_{\pm}\sum_{n\ge0}t_{\pm,n}\,\lamstar^{-\beta+u_{\pm,n}}\,,
\end{equation}
with $u_{\pm,n}=\pm\nueff+\tfrac12-2n$. The integer series is built on the
continued moments $M_a(-n)$, and produces the effective field theory part of the
bispectrum, vanishing identically as $\rhohat\to0$. The two collider series are
built on the tail coefficients $D^{(a)}_{\pm}$ \eqref{eq:tails}, and bring the non-analytic part. The split of the bispectrum into a smooth background and
collider oscillations for $m_{\rm eff} \geq 3H/2$ is directly inherited from the kernel. Each series converges geometrically, uniformly
over the triangle, the equilateral configuration being the slowest point with ratio $1/2$.

Inserting \eqref{eq:towers} into \eqref{eq:master} makes the entire $k_3$
dependence explicit and leaves, term by term, a single object, the
\emph{hard-leg integral},
\begin{equation}
\label{eq:pairoverlap}
\mathcal G_{a_1a_2}(p)=\iint_1^{\infty}\d\lambda_1\d\lambda_2\,
\frac{B_{a_1}(\lambda_1)\,B_{a_2}(\lambda_2)}
{\big(k_1\lambda_1+k_2\lambda_2\big)^{p}}\,.
\end{equation}
Physically, this is the strong-mixing analogue of a familiar
object. In the weak-mixing computation of the collider signal (see references above), the massive
field is exchanged between two vertices, and the two time integrals are
nested. In the squeezed limit they factorize, the soft one fixing the
non-analytic $k_3$-dependence responsible for the oscillation, and the hard one, where the two hard modes meet
the massive line, fixing the amplitude and the phase of the signal. Our strong-mixing calculation proceeds in the same way: the expansion
\eqref{eq:towers} of the soft leg carries the kinematic dependence in the squeezed limit, and
\eqref{eq:pairoverlap} carries amplitude and phase. Here, however, every
interaction is a contact one and nothing is exchanged: each hard leg
carries the massive field through the Laplace kernel.
To compute \eqref{eq:pairoverlap}, we map each frequency to the unit interval,
$\lambda_i=(1+v_i)/(1-v_i)$, under which the kinematics factorizes as
\begin{equation}
\label{eq:kernelid}
k_1\lambda_1+k_2\lambda_2
=k_{12}\,\frac{(1-v_1v_2)+\delta\,(v_1-v_2)}{(1-v_1)(1-v_2)}\,,
\end{equation}
with $k_{12}\equiv k_1+k_2$ and $\delta=(k_1-k_2)/k_{12}$. On the unit square, $|v_1-v_2|\le1-v_1v_2$, while on the triangle $|\delta|\le\tfrac13$, since
$k_1\le k_2+k_3\le 2k_2$. The expansion of \eqref{eq:pairoverlap} in powers of $\delta$ therefore converges
geometrically, uniformly over the triangle, and fastest at
$\delta=0$, the isosceles configurations. Each order then leaves an
integral of Euler type over the unit square, which evaluates to
Beta functions~\cite{NIST:DLMF}, as shown in \SM, completing the analytical
computation of the bispectrum.
The gradient interaction
$(\partial_i\pic)^{2}\sigma$ requires a separate treatment.

\paragraph*{\bf Shapes and collider signal.}
The observationally relevant quantity is the shape of the bispectrum~\cite{Babich:2004gb}, defined such that $\langle\zeta_{\bm k_1}\zeta_{\bm k_2}\zeta_{\bm k_3}\rangle
=(2\pi)^{7}\,\delta^{(3)}(\sum \bm k_i)\,\Delta_\zeta^{4}\,S(k_1,k_2,k_3)/(k_1k_2k_3)^{2}$, with $\Delta_\zeta^{2}=2.1 \times 10^{-9}$ the measured power spectrum amplitude on CMB scales~\cite{Planck:2018jri}.
At strong mixing, the overall amplitude of the shape, as conventionally
measured at the equilateral point, is known to reach large values while remaining under
perturbative control~\cite{Werth:2023pfl,Pinol:2023oux}. This also holds for the self-interaction $\dot\pi_c^{3}$, not covered in those works: in
the regime $\rhohat\gtrsim\mhat^{2}$, dimensional analysis in the non-local
single-field theory obtained by integrating out $\sigma$ gives
$f_{\rm NL}^{\dot\pi_c^{3}}\sim(H^{2}\lambda_2/\Delta_\zeta)\,\rhohat^{-3/4}$,
with perturbative control extending up to $f_{\rm NL}\sim\Delta_\zeta^{-1}$.
We therefore normalize our shapes at the equilateral configuration, and focus on
what the amplitude alone does not capture. The left panel of Fig.~\ref{fig:shapes} shows the shape of the first four cubic interactions in
\eqref{eq:lagrangian} along the isosceles family $S(1,1,\kappa)$, with $\kappa=k_3/k_1$, for a representative theory at strong mixing,
$\mhat^{2}=1.25$ and $\mueff \equiv -i \nueff= 3$. For such parameters, the field
is too light to oscillate on its own, and the oscillations are created by the mixing.
Overlaid are the numerical results obtained with CosmoFlow in~\cite{Philcox:2026tjj},
with excellent agreement between the two methods to relative accuracy $10^{-5}$. Numerical computations however become prohibitive at large squeezing, and
configurations with $\kappa < 0.014$ (dashed line) are handled by a fitted form in~\cite{Philcox:2026tjj}. By contrast, our formulas treat all configurations alike, in about two seconds on a laptop for the four interactions in Fig.~\ref{fig:shapes}, demonstrating the power of analytical methods.

For $\kappa \ll 1$, the collider series of \eqref{eq:towers} reduce to their leading
terms and one gets the cosmological collider oscillations of frequency $\mueff$, $S\simeq\mathcal A\,\kappa^{1/2}\cos\big(\mueff\ln \kappa+\varphi\big)$, whose amplitude and phase are read off the expansion. At large mixing and fixed $\mueff$, the amplitude $\mathcal A$ of the signal was shown to grow like $e^{\pi\rhohat/2}$~\cite{Pinol:2026xnl}. While this can lead to interesting enhancements for moderate values of $\rho$, that regime features a transient tachyonic instability and parametrically lies in a region where perturbative control is lost due to graviton loops~\cite{Garcia-Saenz:2025jis}. It therefore requires further study. Instead, we concentrate here on the conventional strong-mixing regime at fixed $m^2 >0$, varying $\rho \gtrsim m$. Each leg in \eqref{eq:assembly} carries a weight $e^{\pm\pi\rhohat/2}$ according to its channel, so the dominant term, with all three legs in the amplified channel, scales like $e^{+3\pi\rhohat/2}$, for all interactions.
The collider coefficients of the soft leg, $t_{\pm,0}$ in \eqref{eq:towers}, cost $e^{-\pi\mueff}$, and the hard-leg integral \eqref{eq:pairoverlap} a further $e^{-\pi\rhohat}$. Deep in the strong-mixing regime where $ \rho \simeq \mueff $, one thus finds
\begin{equation}
\label{eq:halfboltz}
\mathcal A\;\propto\;e^{-\pi\mueff/2}
\qquad\text{(strong mixing)}\,,
\end{equation}
only half the conventional Boltzmann suppression that had been anticipated by analogy with weak mixing in~\cite{Pinol:2023oux,Philcox:2026tjj}.
As the right panel of Fig.~\ref{fig:shapes} shows, this parametric scaling already holds at the moderate values of $\mueff$ of observational relevance.
The power-law prefactors, by contrast, sharply distinguish the interactions. At
fixed $\mhat^{2}$ we find, more precisely,
\begin{equation}
\label{eq:powerlaw}
\mathcal A\;\propto\;\mueff^{\,n}\,e^{-\pi\mueff/2}\,,
\qquad n=0,\ 0,\ 2,\ 3\,,
\end{equation}
for $\sigma^{3}$, $\dot\pi_c\sigma^{2}$, $\dot\pi_c^{2}\sigma$ and
$\dot\pi_c^{3}$ respectively: one power of $\mueff$ per $\dot\pi_c$ leg.\footnote{The
vertex $\dot\pi_c\sigma^{2}$ escapes the counting: each of the three cyclic
permutations of the momenta separately carries the expected power, which
cancels in their sum.} The origin is
simple. The collider signal is generated when the pair is created, at
$|k_t\tau|\sim\mueff$, with the total energy $k_t=k_1+k_2+k_3$ dominated by the
hard momenta, while the smooth part of the bispectrum forms around the
total-energy horizon crossing, $|k_t\tau|\sim1$. Each $\dot\pi_c$ leg
removes one power of the scale factor and thus weights the vertex integrand by
one extra power of conformal time. Evaluated at pair creation rather than
at horizon crossing, each such power contributes a factor $\mueff$. The smooth parts of the four shapes are instead comparable, so the whole
difference in visibility lies in the oscillations, as the left panel makes
clear: at strong mixing, interactions with more time derivatives are
parametrically the cleanest collider targets. For $\dot\pi_c^{3}$ with the
parameters of Fig.~\ref{fig:shapes}, the shape reaches four times its
equilateral value, so that the equilateral amplitude alone substantially
understates the signal.

\paragraph*{\bf Conclusion.}
We have solved the strongly mixed regime of a massive field coupled to the
curvature perturbation, from the linear theory to the bispectrum. The
construction is systematic: a single kernel carries the whole theory, every
interaction goes through the same chain, and the bispectra come out as
closed-form series of elementary functions. They are evaluated in seconds at any
point in parameter space, and validated from the deep squeezed limit to the
equilateral configuration. Among the recent routes into this regime, the exact linear
solutions of~\cite{Huenupi:2026abj,Huenupi:2026aqc} were first obtained as series of derivatives of Tricomi functions. The exponential
integral representations they later reach, shared by
\cite{Pinol:2026xnl}, follow, after integrations by parts, from the plane-wave superposition
that the Laplace representation derives systematically. At the level of the
bispectrum, both routes lead to exact integral representations, studied in
the squeezed limit in~\cite{Huenupi:2026aqc} and evaluated numerically for the interaction $\dot\pi_c^{3}$ in~\cite{Pinol:2026xnl}, while the results of~\cite{Wang:2026lff} are obtained, in practice, close to the weakly mixed regime.

Analytic control first brings understanding: the halved Boltzmann
suppression \eqref{eq:halfboltz} follows from an explicit accounting of the
exponential factors, and the hierarchy among interactions has an equally
transparent origin, the weight of the vertex at the time of pair creation.
It also provides observational guidance. The collider signal at strong
mixing is both larger than long believed and unevenly distributed: it
concentrates in the interactions with time derivatives of the curvature
perturbation, where the oscillations can exceed the equilateral amplitude
that conventional searches are calibrated on. Analytic shapes are
templates, as accurate deep in the squeezed limit as anywhere else and
evaluable at negligible cost. The searches for the collider signal now under
way~\cite{Sohn:2024xzd,Cabass:2024wob,Philcox:2025bbo,Suman:2025vuf,Kumar:2026ogn,Philcox:2026tjj} would directly benefit from them.

Nothing in this construction, finally, is tied to the model solved here. The
representation asks nothing of the theory: any mode function can be written
as a superposition of flat-space plane waves, with the entire dynamics
condensed into the kernel of that superposition, here an associated Legendre
function. The same strategy now points to spinning fields, modified
dispersion relations, and beyond the bispectrum: at strong mixing every
tree-level bispectrum is a contact diagram, the mixing already resummed
in the mode functions, so the first genuinely new object is the trispectrum.
These are the next steps of the Laplace-space approach to cosmological
correlators.

\paragraph*{\bf Acknowledgements.} We thank Sebastian Garcia-Saenz for useful discussions.

\bibliography{references-Laplace-clean.bib}

\clearpage
\onecolumngrid
\begin{center}
\textbf{\large Supplemental Material}
\end{center}
\setcounter{equation}{0}
\renewcommand{\theequation}{S\arabic{equation}}

\noindent This Supplemental Material is organized as follows. The first section is devoted to the moment function of the kernel, the central object of the calculation: how it is analytically continued outside the convergence domain of its definition \eqref{eq:momentfunction}, in particular to regulate apparently divergent integrals. As a first application, it gives the late-time values $\pi_{c,a}(0)$ used throughout the main text.
The second section carries out the first step of the
bispectrum computation, the expansion of the soft-leg integral \eqref{eq:softblock} into three series, whose
coefficients are provided by the moment function. The third evaluates the hard-leg integral.
The fourth gathers the soft-leg expansion and the hard-leg integrals to give the full form of the bispectra.

\section*{The moment function of the kernel}

One object underlies every stage of this calculation: the \emph{moment
function} of the kernel, introduced in Eq.~\eqref{eq:momentfunction} of the
main text,
\begin{equation}
\label{eq:smMdef}
M_a(u)=\int_1^{\infty}\!\d\lambda\;\lambda^{-u}\,B_a(\lambda)\,.
\end{equation}
Equation~\eqref{eq:smMdef} is an integral depending on a parameter: it defines
an analytic function of $u$ on its domain of convergence, extended elsewhere by
analytic continuation, exactly as the Euler integral defines the Beta function.
It is the Mellin transform of the kernel restricted to the cut, and the
correspondence between its poles and the behavior of $B_a$ at large $\lambda$
is the standard dictionary of Mellin analysis. When $\nueff$ is imaginary, the two tails
\eqref{eq:tails} decay as
$\lambda^{-1/2}$, so \eqref{eq:smMdef} converges
only for $\mathrm{Re}\,u>\tfrac12$. All the values needed below lie outside that range, and the continuation is genuinely needed.

\paragraph*{Finite parts.} For any exponent $w$,
\begin{equation}
\label{eq:smincgamma}
\int_1^{\infty}\!\d\lambda\;\lambda^{w}\,e^{-i\lambda z}
=(iz)^{-1-w}\,\Ga(1+w,iz)\,,
\end{equation}
with $\Ga(\cdot,\cdot)$ the upper incomplete Gamma function. Expanding the
latter around $z=0$ gives
\begin{equation}
\label{eq:smsplit}
\int_1^{\infty}\!\d\lambda\;\lambda^{w}\,e^{-i\lambda z}
=\Ga(1+w)\,(iz)^{-1-w}
\;-\;\sum_{n\ge0}\frac{(-iz)^{n}}{n!\,(w+n+1)}\,.
\end{equation}
The constant term of the series is $-1/(w+1)$, which is what
$\int_1^{\infty}\lambda^{w}\d\lambda$ would give if one continued it in $w$.
Nothing here is a prescription imposed on a divergent integral: the left-hand
side is finite for every $z$ with $\Im(z)<0$, and \eqref{eq:smsplit} is its expansion.
Now let $f$ be any function whose non-integrable behavior at infinity is a
finite sum of powers, $f(\lambda)\simeq\sum_i c_i\,\lambda^{w_i}$. Applying
\eqref{eq:smsplit} to each power and letting $z\to0$ on the convergent
remainder gives
\begin{equation}
\label{eq:smfp}
\lim_{z\to0}\Bigg[\int_1^{\infty}\!\d\lambda\;f(\lambda)\,e^{-i\lambda z}
-\sum_i c_i\,\Ga(1+w_i)\,(iz)^{-1-w_i}\Bigg]
=\int_1^{\infty}\!\d\lambda\,\Big[f(\lambda)-\sum_i c_i\,\lambda^{w_i}\Big]
-\sum_i\frac{c_i}{w_i+1}\,.
\end{equation}
The left-hand side regulates the integral with a plane wave and strips off its
fractional powers.
The right-hand side subtracts the non-integrable powers, integrates the convergent remainder, and adds the powers back with
their integrals analytically continued in the exponent.
Equation~\eqref{eq:smfp} says that the two coincide; their common value is the \emph{finite
part} of the integral. Every divergent-looking integral met in this Letter is read this way: the
regulator is the plane wave the kernel is integrated against, and the
subtracted behaviors are the exact tails of the kernel.

The behaviors to subtract are read off the form of the kernel adapted to large
$\lambda$:
\begin{equation}
\label{eq:smsector}
B_a(\lambda)=\sum_{\sigma=\pm}D^{(a)}_{\sigma}\;
\lambda^{\sigma\nueff-\frac12}\,\big(1-\lambda^{-2}\big)^{-s}\;
{}_2F_1\!\left[\genfrac{}{}{0pt}{1}
{\alpha_\sigma,\ \alpha_\sigma+\tfrac12}{1-\sigma\nueff};\lambda^{-2}\right],
\qquad \alpha_\sigma=\tfrac14-\tfrac{\sigma\nueff}{2}-s\,.
\end{equation}
Expanding \eqref{eq:smsector} in powers of $\lambda^{-2}$ gives
\begin{equation}
    B_a(\lambda)=\sum_{\sigma=\pm} \sum_{n=0}^\infty R^{(a)}_{\sigma,n} \,\lambda^{-\frac12+\sigma\nueff-2n}\,,
\end{equation}
with the leading coefficients being the tails themselves, $R^{(a)}_{\pm,0}=D^{(a)}_{\pm}$ of \eqref{eq:tails}, and the higher ones
terminating sums built on them,
\begin{equation}
\label{eq:smres}
\boxed{\;
R^{(a)}_{\pm,n}=D^{(a)}_{\pm}\;\frac{(-1)^{n}}{n!}\,
\frac{\Ga(1-s)}{\Ga(1-n-s)}\;
\FF{3}{2}\!\left[\genfrac{}{}{0pt}{1}
{\alpha_{\pm},\ \alpha_{\pm}+\tfrac12,\ -n}
{1\mp\nueff,\ 1-n-s};1\right],\;}
\end{equation}
each a sum of $n+1$ terms. The moment function is then defined, for any
large-enough integer $J$, by
\begin{equation}
\label{eq:smMfp}
M_a(u)=\int_1^{\infty}\!\d\lambda\;\lambda^{-u}
\Big[B_a(\lambda)-\sum_{\pm}\sum_{n<J}R^{(a)}_{\pm,n}\,
\lambda^{-\frac12\pm\nueff-2n}\Big]
\;+\;\sum_{\pm}\sum_{n<J}\frac{R^{(a)}_{\pm,n}}{u-u_{\pm,n}}\,,
\end{equation}
with $u_{\pm,n}=\pm\nueff+\tfrac12-2n$ as in the main text.
Equation~\eqref{eq:smMfp} displays the whole
analytic structure: $M_a$ is meromorphic in $u$, with two sets of simple
poles at $u=u_{\pm,n}$ whose residues are the $R^{(a)}_{\pm,n}$.

\paragraph*{Closed forms.} The moment function follows from
\eqref{eq:smsector} term by term. Under $x=\lambda^{-2}$ each power
is a Beta integral,
\begin{equation}
\label{eq:smsectorint}
\int_1^{\infty}\!\d\lambda\;\lambda^{-u+\sigma\nueff-\frac12-2m}
\big(1-\lambda^{-2}\big)^{-s}
=\frac12\,\mathrm{B}\big(y_\sigma+m,\,1-s\big)\,,
\qquad
y_\sigma=\frac{u-\sigma\nueff-\tfrac12}{2}\,,
\end{equation}
convergent for $\mathrm{Re}\,y_\sigma+m>0$ and continued otherwise. With
$\mathrm{B}(y+m,c)=\mathrm{B}(y,c)\,(y)_m/(y+c)_m$, one then obtains
\begin{equation}
\label{eq:smMu}
M_a(u)=\frac12\sum_{\sigma=\pm}D^{(a)}_{\sigma}\;
\mathrm{B}\big(y_\sigma,\,1-s\big)\,
\FF{3}{2}\!\left[\genfrac{}{}{0pt}{1}
{\alpha_\sigma,\ \alpha_\sigma+\tfrac12,\ y_\sigma}
{1-\sigma\nueff,\ y_\sigma+1-s};1\right].
\end{equation}
Its poles are again the same two sets, produced here by the Beta functions
$\mathrm{B}(y_\sigma,1-s)$. In practice, we need to evaluate $M_a$ only at
non-positive integers, where the results reduce to Gamma functions alone.
Indeed, the parameters of \eqref{eq:smMu} obey the exact relation
$y_\sigma+1-s=\alpha_\sigma+\tfrac12+\tfrac{u}{2}$: at $u=0$ the lower
parameter coincides with the upper parameter $\alpha_\sigma+\tfrac12$, at
$u=-1$ with $\alpha_\sigma$, and in both cases the $_3F_2$
reduces to a Gauss series at unit argument, summed in closed
form~\cite[eq.~(15.4.20)]{NIST:DLMF}. At $u=0$,
\begin{equation}
\label{eq:smgausszero}
\FF{3}{2}\!\left[\genfrac{}{}{0pt}{1}
{\alpha_\sigma,\ \alpha_\sigma+\tfrac12,\ y_\sigma}
{1-\sigma\nueff,\ y_\sigma+1-s};1\right]_{u=0}
={}_2F_1\big(\alpha_\sigma,\,y_\sigma;\,1-\sigma\nueff;\,1\big)
=\frac{\Ga(1-\sigma\nueff)\,\Ga(1+s)}
{\Ga\!\big(\tfrac34-\tfrac{\sigma\nueff}{2}+s\big)\,
\Ga\!\big(\tfrac54-\tfrac{\sigma\nueff}{2}\big)}\,.
\end{equation}
The two sectors then combine, by the reflection and duplication formulas of
the Gamma function, into
\begin{equation}
\label{eq:smA0closed}
M_a(0)=\frac{-s\;2^{\,2s+1}\;\Ga(1-2s)\;
\Ga\!\big(\tfrac34-\tfrac{\nueff}{2}\big)\Ga\!\big(\tfrac34+\tfrac{\nueff}{2}\big)}
{\big(\tfrac14-\nueff^{2}\big)\;
\Ga\!\big(\tfrac34-\tfrac{\nueff}{2}-s\big)\Ga\!\big(\tfrac34+\tfrac{\nueff}{2}-s\big)}
\end{equation}
and the same two steps at $u=-1$ give
\begin{equation}
\label{eq:smA1closed}
\;M_a(-1)=\frac{-s\;2^{\,2s+1}\;\Ga(1-2s)\;
\Ga\!\big(\tfrac14-\tfrac{\nueff}{2}\big)\Ga\!\big(\tfrac14+\tfrac{\nueff}{2}\big)}
{\big(\mhat^{2}+\rhohat^{2}\big)\;
\Ga\!\big(\tfrac14-\tfrac{\nueff}{2}-s\big)\Ga\!\big(\tfrac14+\tfrac{\nueff}{2}-s\big)}\,.
\end{equation}
The other moments are treated similarly. At $u=-n$, with
$n=2m+p$ and $p\in\{0,1\}$, the lower parameter reads
$\alpha_\sigma+\tfrac{p}{2}-m$, an upper parameter shifted down by an
integer. The ratio of their Pochhammer symbols is then a polynomial in the
summation index, the series truncates onto Gauss sums, and one finds that every continued
moment reads
\begin{equation}
\label{eq:smMn}
\boxed{\;\begin{aligned}
M_a(-n)={}&\tfrac12\,\Ga(1-s)\,\Ga(1+s+m)\\[2pt]
&\times\sum_{\sigma=\pm}D^{(a)}_{\sigma}\,
\frac{\Ga(y_\sigma)\,\Ga(1-\sigma\nueff)}
{\Ga(y_\sigma+1-s)\,
\Ga\!\big(\tfrac34-\tfrac{\sigma\nueff}{2}+s-\tfrac{p}{2}\big)\,
\Ga\!\big(\tfrac54+\tfrac{n}{2}-\tfrac{\sigma\nueff}{2}\big)}\,
\FF{3}{2}\!\left[\genfrac{}{}{0pt}{1}
{-m,\ \alpha_\sigma+\tfrac{p}{2},\ y_\sigma}
{y_\sigma+1-s,\ -s-m};1\right]
\end{aligned}\;}
\end{equation}
with $y_\sigma=-\tfrac{n}{2}-\tfrac14-\tfrac{\sigma\nueff}{2}$ its value at
$u=-n$, and with the $_3F_2$ a terminating sum of $m+1$ terms.
The cases $m=0$ reproduce \eqref{eq:smA0closed} and \eqref{eq:smA1closed}.
Notice that integrating the dual operator $\mathcal L_2$ of Eq.~\eqref{eq:L2} by parts,
with no boundary terms, gives the recursion relation
\begin{equation}
\label{eq:smrec}
\big[(u-2)(u-3)+\mhat^{2}+\rhohat^{2}-2\big]M_a(u-2)
-\big[2(u-1)^{2}+\mhat^{2}-2\big]M_a(u)
+u(u+1)\,M_a(u+2)=0\,.
\end{equation}
Its coefficient $u(u+1)$ vanishes at both $u=0$ and $u=-1$, and one can check that \eqref{eq:smMn} follows from the two values \eqref{eq:smA0closed} and \eqref{eq:smA1closed} alone.

\paragraph*{The late-time values.} The late-time values $\pi_{c,a}(0)$ now follow.
We denote by $\mathcal P(\lambda,z)$ the bracket in
\eqref{eq:picray} and split the kernel as in \eqref{eq:smMfp}:
\begin{equation}
\label{eq:smfvsplit}
\pi_{c,a}(z)=-\frac{\sqrt{w_a}}{\rhohat}\Bigg\{
\int_1^{\infty}\!\d\lambda\,\Big[B_a-\!\!\sum_{\pm,\,n<J}\!\!
R^{(a)}_{\pm,n}\,\lambda^{-\frac12\pm\nueff-2n}\Big]\,
e^{-i\lambda z}\,\mathcal P
\;+\;\sum_{\pm,\,n<J}R^{(a)}_{\pm,n}\!\int_1^{\infty}\!\d\lambda\;
\lambda^{-\frac12\pm\nueff-2n}\,e^{-i\lambda z}\,\mathcal P\Bigg\}\,.
\end{equation}
For $J$ large enough the limit can be taken under the first integral, where only the
constant term of $\mathcal P$ survives,
\begin{equation}
\label{eq:smfvrem}
\int_1^{\infty}\!\d\lambda\,\big[B_a-\cdots\big]\,e^{-i\lambda z}\,\mathcal P
\;\xrightarrow[\ z\to0\ ]{}\;
-\big(\mhat^{2}+\rhohat^{2}-2\big)\int_1^{\infty}\!\d\lambda\,
\big[B_a-\cdots\big]\,,
\end{equation}
while in the second term of \eqref{eq:smfvsplit}, every monomial integral is elementary by
\eqref{eq:smincgamma}: with $w=-\tfrac12\pm\nueff-2n$,
\begin{equation}
\label{eq:smfvtail}
\int_1^{\infty}\!\d\lambda\;\lambda^{w}\,e^{-i\lambda z}\,\mathcal P(\lambda,z)
=\big(\text{fractional powers of }z\big)
\;+\;\frac{\mhat^{2}+\rhohat^{2}-2}{w+1}\;+\;O(z)\,.
\end{equation}
With $w+1=u_{\pm,n}$, \eqref{eq:smfvrem} and \eqref{eq:smfvtail}
reassemble into the finite part
\eqref{eq:smMfp} at $u=0$,
\begin{equation}
\label{eq:smfrozen}
\boxed{\;
\pi_{c,a}(0)=\sqrt{w_a}\;\frac{\mhat^{2}+\rhohat^{2}-2}{\rhohat}\,M_a(0)
=- s\frac{\sqrt{w_a}}{\rhohat} \frac{2^{\,2s+1}\;\Ga(1-2s)\;
\Ga\!\big(\tfrac34-\tfrac{\nueff}{2}\big)\Ga\!\big(\tfrac34+\tfrac{\nueff}{2}\big)}
{\Ga\!\big(\tfrac34-\tfrac{\nueff}{2}-s\big)\Ga\!\big(\tfrac34+\tfrac{\nueff}{2}-s\big)}\,.\;}
\end{equation}
One can check explicitly that the ratio of these
late-time values satisfies \eqref{eq:asymmetry}. With $\lvert\Ga(1-2s)\rvert^{2}=\pi\rhohat/\sinh(\pi\rhohat)$ and the explicit expression \eqref{eq:weights-w} of $w_a$, the power spectrum reads $\Delta_\zeta^{2}=R\,\Delta_{\zeta,0}^{2}$, where $\Delta_{\zeta,0}^{2}=(4 \pi^2)^{-1}(H/f_\pi)^4$ is the uncoupled value and the amplification factor $R \geq 1$ is given by
\begin{equation}
\label{eq:smR}
R=\left|\frac{\Ga\!\big(\tfrac34-\tfrac{\nueff}{2}\big)
\Ga\!\big(\tfrac34+\tfrac{\nueff}{2}\big)}
{\Ga\!\big(\tfrac34-\tfrac{\nueff}{2}+\tfrac{i\rhohat}{2}\big)
\Ga\!\big(\tfrac34+\tfrac{\nueff}{2}+\tfrac{i\rhohat}{2}\big)}\right|^{2}\,,
\end{equation}
obtained here independently, in agreement with~\cite{Huenupi:2026abj,Pinol:2026xnl}.

\section*{Expansion of the soft-leg integral}

We order $k_3\le k_2\le k_1$. Whatever the interaction, the Laplace frequency of
the softest leg enters the vertex integral only through
\begin{equation}
\label{eq:smT}
T_\beta(\lamstar)=\int_1^\infty\!\d\lambda\;\frac{B_a(\lambda)}{(\lambda+\lamstar)^{\beta}}\,,
\qquad \lamstar=\frac{k_1\lambda_1+k_2\lambda_2}{k_3}\,,
\end{equation}
where $\lamstar$ is the Laplace frequency at which the soft leg sees the hard
pair, the total energy being $K=k_3(\lambda_3+\lamstar)$. One has $\beta=3$ for
$\sigma^{3}$, shifted by integers for the other interactions. Since $k_3$ is
the smallest momentum and every $\lambda_i\ge1$, one always has $\lamstar \geq 2$. The denominator has an exact Mellin--Barnes representation, the inverse
Mellin transform of the Beta integral~\cite[eq.~(5.12.3)]{NIST:DLMF},
\begin{equation}
\label{eq:smMBker}
\frac{1}{(\lambda+\lamstar)^{\beta}}
=\frac{1}{2\pi i}\int_{c-i\infty}^{c+i\infty}\!\d t\;
\frac{\Ga(t)\,\Ga(\beta-t)}{\Ga(\beta)}\;\lambda^{\,t-\beta}\,\lamstar^{-t}\,,
\qquad 0<c<\mathrm{Re}\,\beta\,,
\end{equation}
and integrating the kernel against $\lambda^{\,t-\beta}$ produces the moment
function,
\begin{equation}
\label{eq:smMB}
T_\beta(\lamstar)=\frac{1}{2\pi i}\int_{c-i\infty}^{c+i\infty}\!\d t\;
\frac{\Ga(t)\,\Ga(\beta-t)}{\Ga(\beta)}\;M_a(\beta-t)\;\lamstar^{-t}\,,
\qquad 0<c<\mathrm{Re}\,\beta-\tfrac12\,,
\end{equation}
the tighter condition ensuring that the $\lambda$ integral converges on the
contour, $\mathrm{Re}\,(\beta-t)>\tfrac12$. The strip separates the poles of
$\Ga(t)$, at $t=-n$, on its left, from everything else on its right: the
poles of $\Ga(\beta-t)$, at $t=\beta+n$, and the two sets of poles of
$M_a$ displayed in \eqref{eq:smMfp}, at $t=\beta-u_{\pm,n}$, whose real
parts $\mathrm{Re}\,\beta-\tfrac12+2n$ all exceed $c$. Since
$\lamstar>1$, the factor $\lamstar^{-t}$ decays to the right and the contour
closes there, giving
\begin{equation}
\label{eq:smtowers}
\boxed{\;T_\beta(\lamstar)
=\sum_{n\ge0}t^{(\beta,a)}_{n}\;\lamstar^{-\beta-n}
+\sum_{\pm}\sum_{n\ge0}t^{(\beta,a)}_{\pm,n}\;\lamstar^{-\beta+u_{\pm,n}}\,,\;}
\end{equation}
with coefficients
\begin{equation}
\label{eq:smtcoefs}
t^{(\beta,a)}_{n}=\frac{(-1)^{n}}{n!}\,\frac{\Ga(\beta+n)}{\Ga(\beta)}\,M_a(-n)\,,
\qquad
t^{(\beta,a)}_{\pm,n}=R^{(a)}_{\pm,n}\,
\frac{\Ga(\beta-u_{\pm,n})\,\Ga(u_{\pm,n})}{\Ga(\beta)}\,.
\end{equation}
These are the coefficients $t_{n}$ and $t_{\pm,n}$ of Eq.~\eqref{eq:towers}
of the main text, with their dependence on $\beta$ and on the channel
shown here explicitly. From \eqref{eq:smMn} and \eqref{eq:smres}, both sets are given in terms of
$\Gamma$-functions and terminating sums.
Each series converges geometrically, uniformly
over the triangle, the equilateral configuration being the slowest point with ratio $1/2$.

The integer series is the analytic part of the soft leg. Its coefficients
vanish as $\rhohat\to0$, and it builds the smooth background of the shape. The
other two carry the non-analytic powers $\lamstar^{1/2\pm \nueff}$, that is, the
collider oscillations for $\nueff^2 <0$, with amplitudes set by the tails \eqref{eq:tails}. At $\beta=1$, the two collider series resum to
$\Ga(\tfrac12+\nueff)\Ga(\tfrac12-\nueff)B_a(\lamstar)$, so that the transform
returns the kernel itself, up to the integer series. As $\rhohat\to0$, the
integer series vanishes and the collider series resum to the known result for a
free massive field~\cite{Belrhali:2026ygh,Belrhali:2026jqe}.

\section*{Closed form of the hard-leg integral}

Inserting the series \eqref{eq:towers} into the vertex integral
\eqref{eq:master} leaves, term by term, the hard-leg integrals
$\mathcal G_{a_1a_2}(p)$ of Eq.~\eqref{eq:pairoverlap}. The exponents needed
are $p=\beta+n$ and $p=\beta-u_{\pm,n}$, with $\beta=3$ for \eqref{eq:master}
and shifted by integers for the other interactions. We keep $\beta$ general.
We use the change of variable $\lambda=(1+v)/(1-v)$, which maps each
Laplace-frequency half-line to the unit interval: the Bunch--Davies endpoint
$\lambda=1$ goes to $v=0$, and the far tails $\lambda\to\infty$ to $v=1$. Equation~\eqref{eq:pairoverlap} then becomes
\begin{equation}
\label{eq:pairoverlap_01}
\mathcal G_{a_1a_2}(p)
=\frac{4}{k_{12}^p}\int_0^1\int_0^1\d v_1\d v_2\,
B_{a_1}\left(\frac{1+v_1}{1-v_1}\right)\, B_{a_2}\left(\frac{1+v_2}{1-v_2}\right)
\left[(1-v_1)(1-v_2)\right]^{p-2}\left[(1-v_1v_2)+\delta\,(v_1-v_2)\right]^{-p}
\,,
\end{equation}
with $k_{12}=k_1+k_2$ and $\delta=(k_1-k_2)/k_{12}$, as in \eqref{eq:kernelid}.
In this variable the Legendre kernel \eqref{eq:kernel} becomes a Gauss
function~\cite[eq.~(14.3.6)]{NIST:DLMF},
\begin{equation}
\label{eq:smFa}
B_a\!\left(\frac{1+v}{1-v}\right)=v^{-s}\,(1-v)^{\frac12-\nueff}F_a(v)\,,\qquad
F_a(v)={}_2F_1\!\left[\genfrac{}{}{0pt}{1}
{\tfrac12-\nueff,\ \tfrac12-\nueff-2s}{1-2s};v\right]\,.
\end{equation}
The power $v^{-s}$ is
the dual image of the early-time phases \eqref{eq:earlytime}. At $v=1$ the
explicit prefactor $(1-v)^{\frac12-\nueff}$ carries the first tail of
\eqref{eq:tails}, while $F_a$ is itself non-analytic there, with the two local
behaviors $(1-v)^{0}$ and $(1-v)^{2\nueff}$. The kernel as a whole therefore behaves as
$(1-v)^{\frac12\mp\nueff}$. The defining series of \eqref{eq:smFa} converges
about $v=0$, the Bunch--Davies endpoint. The connection formula of the Gauss
function~\cite[eq.~(15.8.4)]{NIST:DLMF} re-expands the same function about
$v=1$ instead, giving
\begin{equation}
\label{eq:smconnection}
F_a(v)=2^{-\frac12+\nueff}D^{(a)}_{+}\,
\FF{2}{1}\!\left[\genfrac{}{}{0pt}{1}
{\tfrac12-\nueff,\ \tfrac12-\nueff-2s}{1-2\nueff};1-v\right]
\;+\;2^{-\frac12-\nueff}D^{(a)}_{-}\,(1-v)^{2\nueff}\,
\FF{2}{1}\!\left[\genfrac{}{}{0pt}{1}
{\tfrac12+\nueff,\ \tfrac12+\nueff-2s}{1+2\nueff};1-v\right]\,.
\end{equation}
The first term is analytic at $v=1$, the second carries the singular branch,
and expanding its $\FF{2}{1}$ in powers of $1-v$ gives the singular orders one
by one in closed form. We build this into the calculation from the
start, by writing, for any integer $J\ge0$,
\begin{equation}
\label{eq:smdecomp}
F_a(v)=\sum_{\ell=0}^{J-1}h^{(a)}_{\ell}\,(1-v)^{2\nueff+\ell}
\;+\;F^{(a,J)}(v)\,,\qquad
h^{(a)}_{\ell}=2^{-\frac12-\nueff}D^{(a)}_{-}\;
\frac{\big(\tfrac12+\nueff\big)_{\ell}\,\big(\tfrac12+\nueff-2s\big)_{\ell}}
{(1+2\nueff)_{\ell}\;\ell!}\,,
\end{equation}
which separates the first $J$ singular orders from a remainder, with the
convention that the sum is empty at $J=0$, where $F^{(a,0)}=F_a$.
The remainder has explicit Taylor coefficients at $v=0$,
\begin{equation}
\label{eq:smfJ}
F^{(a,J)}(v)=\sum_{m\ge0}f^{(a,J)}_{m}\,v^{m}\,,\qquad
f^{(a,J)}_{m}=\frac{\big(\tfrac12-\nueff\big)_m\big(\tfrac12-\nueff-2s\big)_m}
{(1-2s)_m\;m!}
-\sum_{\ell=0}^{J-1}h^{(a)}_{\ell}\;\frac{(-2\nueff-\ell)_m}{m!}\,,
\end{equation}
where the second term uses the binomial series
$(1-v)^{2\nueff+\ell}=\sum_{m\ge0}\frac{(-2\nueff-\ell)_m}{m!}\,v^{m}$. These
coefficients decay as $m^{-1-2\nueff-J}$, since the singularity of
$F^{(a,J)}$ is the first one not made explicit in
\eqref{eq:smdecomp}. For imaginary $\nueff$ this is
$m^{-1-J}$ in modulus. The series with $J=0$
decays only as $1/m$, and it is the subtraction that makes the expansion
usable.

Inserting \eqref{eq:smFa} for both legs, the hard-leg integral \eqref{eq:pairoverlap_01} becomes
\begin{equation}
\label{eq:smsquare}
\mathcal G_{a_1a_2}(p)=\frac{4}{k_{12}^{\,p}}
\iint_{(0,1)^2}\prod_{i=1,2}
\Big[v_i^{-s_i}\,(1-v_i)^{X_i-1}\,(1+v_i)^{j_i}\,F_{a_i}(v_i)\Big]\,
\big[(1-v_1v_2)+\delta\,(v_1-v_2)\big]^{-p}\,\d v_1\d v_2\,,
\end{equation}
with $s_i=s(a_i)$ and $X_i=p-j_i-\tfrac12-\nueff$, where $j_i$ is the degree of
the polynomial the leg carries in its Laplace frequency: $j_i=0$ for a $\sigma$
leg, and the finite factor $(1+v_i)^{j_i}$ appears for the derivative legs of
the other interactions. The integrand is singular only at two corners of the
square, $(0,0)$ and $(1,1)$. The kinematic factor is expanded in $\delta$,
\begin{equation}
\label{eq:smdelta}
\big[(1-v_1v_2)+\delta\,(v_1-v_2)\big]^{-p}
=(1-v_1v_2)^{-p}\sum_{r\ge0}\frac{(p)_r}{r!}\,(-\delta)^{r}
\left[\frac{v_1-v_2}{1-v_1v_2}\right]^{r},
\end{equation}
which converges uniformly on the square, since $|v_1-v_2|\le1-v_1v_2$ there
while $|\delta|\le\tfrac13$ on the triangle. Each power is then written in
the variables that vanish at $v_i=1$,
\begin{equation}
\label{eq:smcornervar}
(v_1-v_2)^{r}=\sum_{q=0}^{r}\binom{r}{q}(-1)^{q}\,
(1-v_1)^{q}\,(1-v_2)^{r-q}\,.
\end{equation}
Every term of \eqref{eq:smdelta} thus carries positive powers of $1-v_i$.
Order by order the expansion vanishes at the corner $(1,1)$, and these zeros
exactly compensate the increasingly singular factor $(1-v_1v_2)^{-p-r}$, as the
convergence condition below also shows.

With \eqref{eq:smdecomp}, \eqref{eq:smfJ} and
the finite binomial $(1+v)^{j}=\sum_{k=0}^{j}\binom{j}{k}v^{k}$, each hard leg
is a list of pure powers with closed coefficients,
\begin{equation}
\label{eq:smlegterms}
(1+v)^{j}\,F_a(v)=\sum_{k=0}^{j}\binom{j}{k}
\Bigg[\;\sum_{\ell=0}^{J-1}h^{(a)}_{\ell}\;v^{k}\,(1-v)^{2\nueff+\ell}
\;+\;\sum_{m\ge0}f^{(a,J)}_{m}\;v^{k+m}\Bigg]\,.
\end{equation}
We now define the following convenient notation for the sums in \eqref{eq:smlegterms}:
\begin{equation}
    \sum_{T_i} \ldots\equiv\sum_{k=0}^{j_i} \left[\sum_{\ell=0}^{J-1}\ldots+\sum_{m\geq 0}\ldots \right]\,,
\end{equation}
and for each leg $i=1,2$, we define the $g_i$, $\alpha_i$ and $\gamma_i$ as
\begin{equation}
\label{eq:smdict}
\big(g_i,\ \alpha_i,\ \gamma_i\big)=
\begin{cases}
\Big(\tbinom{j_i}{k}\,f^{(a_i,J)}_{m},\ \ 1-s_i+k+m,\ \ 0\Big)\,,
& k=0,\dots,j_i\,,\quad m\ge0\,,\\[6pt]
\Big(\tbinom{j_i}{k}\,h^{(a_i)}_{\ell},\ \ 1-s_i+k,\ \ 2\nueff+\ell\Big)\,,
& k=0,\dots,j_i\,,\quad \ell=0,\dots,J-1\,,
\end{cases}
\end{equation}
At a given $i$, for each of the two terms of \eqref{eq:smlegterms}, $g_i$, $\alpha_i-1$ and $\gamma_i$ correspond respectively to the remaining coefficients in the sum, the exponent of $v_i$ and that of $1-v_i$, so that the whole leg factor of \eqref{eq:smsquare} can be rewritten compactly as
\begin{equation}
    v_i^{-s_i}\,(1+v_i)^{j_i}\,F_{a_i}(v_i)=\sum_{T_i} \,g_i \,v_i^{\alpha_i-1}\, (1-v_i)^{\gamma_i}\,.
    \label{eq:smlegterms_compact}
\end{equation}
Inserting \eqref{eq:smlegterms_compact} for both legs into \eqref{eq:smsquare}, together with
\eqref{eq:smdelta} and \eqref{eq:smcornervar}, the whole $v$ dependence is a product of
pure powers, and

\begin{equation}
\label{eq:smGasI}
\mathcal G_{a_1a_2}(p)=\frac{4}{k_{12}^{\,p}}
\sum_{r\ge0}\frac{(p)_r}{r!}\,(-\delta)^{r}
\sum_{q=0}^{r}\binom{r}{q}(-1)^{q}
\sum_{T_1,\,T_2}g_1g_2\;
\mathcal I\big(\alpha_1,\alpha_2,c_1,c_2;\,p+r\big)\,,
\end{equation}
the only integral left being
\begin{equation}
\label{eq:smcore}
\mathcal I\big(\alpha_1,\alpha_2,c_1,c_2;\,\sigma\big)=\iint_{(0,1)^2}
v_1^{\alpha_1-1}v_2^{\alpha_2-1}(1-v_1)^{c_1-1}(1-v_2)^{c_2-1}
(1-v_1v_2)^{-\sigma}\,\d v_1\d v_2\,,
\end{equation}
with $c_1=X_1+q+\gamma_1$ and $c_2=X_2+(r-q)+\gamma_2$ the exponents collected
on each variable. The two variables are coupled only through
$(1-v_1v_2)^{-\sigma}$. Expanding that factor in its binomial series decouples
them and leaves Beta functions,
\begin{equation}
\label{eq:smcoreproof}
(1-v_1v_2)^{-\sigma}=\sum_{n\ge0}\frac{(\sigma)_n}{n!}\,v_1^{n}v_2^{n}
\quad\Longrightarrow\quad
\mathcal I=\sum_{n\ge0}\frac{(\sigma)_n}{n!}\,
\mathrm{B}(\alpha_1+n,c_1)\,\mathrm{B}(\alpha_2+n,c_2)\,,
\end{equation}
so that the hard-leg integral finally reads
\begin{equation}
\label{eq:smmaster}
\boxed{\;\mathcal G_{a_1a_2}(p)=\frac{4}{k_{12}^{\,p}}
\sum_{r\ge0}\frac{(p)_r}{r!}\,(-\delta)^{r}
\sum_{q=0}^{r}\binom{r}{q}(-1)^{q}
\sum_{T_1,\,T_2}g_1g_2
\sum_{n\ge0}\frac{(p+r)_n}{n!}\,
\mathrm{B}(\alpha_1+n,c_1)\,\mathrm{B}(\alpha_2+n,c_2)\;}
\end{equation}
where every ingredient is an elementary function. (The sum over $n$ can be
rewritten in terms of a $_3F_2$ at unit argument, but this is not needed.)
The condition for the convergence of the series over $n$ is that
$c_1+c_2-(p+r)$ be positive. This combination equals
$p-j_1-j_2-1-2\nueff+\gamma_1+\gamma_2$, the same for every $r$, $q$, $k$ and
$m$. The singular terms carry $\gamma_i=2\nueff+\ell$ and the regular ones
$\gamma_i=0$, so the smallest value is attained for the regular terms. This
is also the condition for \eqref{eq:pairoverlap} itself to converge, since at
large $\lambda$ the two kernels decay as $\lambda_i^{-\frac12}$ and the double
integral behaves as $\int^{\infty}\!\d\Lambda\;\Lambda^{\,j_1+j_2-p}$. The
series converges wherever the integral it represents is defined.
The interaction fixes both the exponent $p$, reached through the soft-leg
series, and the degrees $j_1,j_2$ of the hard legs. For $\sigma^{3}$,
$\dot\pi_c\sigma^{2}$, $\dot\pi_c^{2}\sigma$ and $\dot\pi_c^{3}$, the smallest
value of this combination is the same, $\tfrac32-3\,\mathrm{Re}\,\nueff$,
hence $\tfrac32$ whenever $\nueff$ is imaginary. For real $\nueff$, that is, for a field
light enough that $\mhat^{2}+\rhohat^{2}<\tfrac94$, this ceases to hold at $\mhat^{2}+\rhohat^{2}=2$. In practice, we used $J=6$ and kept the first seventy terms of
\eqref{eq:smfJ} to produce the results in the main text. This truncation
affects the shapes at the $10^{-6}$ level.

\section*{Full bispectra}

We now assemble all the pieces to build the bispectra.

\paragraph*{The legs.} After the rotation $\tau=it$ to Euclidean time, the leg
of momentum $k$ attached to the vertex depends on time only through $p=kt$. A
$\sigma$ leg is read off \eqref{eq:sigmaray}, a $\dot\pi_c$ leg off the time
derivative of \eqref{eq:picray}, exactly as $\pic$ was obtained from $\sigma$:
\begin{align}
\label{eq:smlegsigma}
\ell^{\sigma}_a(p)&=-\,p^{2}\int_1^{\infty}\!\d\lambda\;B_a(\lambda)\,
e^{-p\lambda}\,,\\
\label{eq:smlegdot}
\ell^{\dot\pi}_a(p)&=\frac{i}{\rhohat}\int_1^{\infty}\!\d\lambda\;B_a(\lambda)
\Big[p^{3}(1-\lambda^{2})+2p^{2}\lambda-(\mhat^{2}-2)\,p\Big]e^{-p\lambda}\,,
\end{align}
each $\dot\pi_c$ leg carrying in addition one explicit factor of its momentum
$k$, from $\dot\pi_c=k\,\pi_c'/a$. Both legs are short lists of monomials
$\tilde c\,p^{m}\lambda^{j}$ integrated against the same kernel: $(m,j,\tilde c)=(2,0,-1)$
for $\sigma$, and $\{(3,0,1),\,(3,2,-1),\,(2,1,2),\,(1,0,-(\mhat^{2}-2))\}$
for $\dot\pi_c$.

\paragraph*{The vertex.} With $n$ the number of $\dot\pi_c$ legs, each removing
one power of $a$, the measure is $a^{4-n}\,\d\tau$, which the rotation
$\tau=it$ turns into $(-1)^{n}\,i^{\,n-1}\,H^{\,n-4}\,\d t/t^{4-n}$.
Combining this with the Wick contractions and the couplings of
\eqref{eq:lagrangian}, one obtains
\begin{equation}
\label{eq:smprefactors}
\big\langle\pic^{3}\big\rangle'
=\mathcal C_v\,(-1)^{n} H^{\,n-4}\,
\mathrm{Im}\Big[\,i^{\,n-1}\sum_{\vec a}
\prod_{i=1}^{3}\sqrt{w_{a_i}}\,\pi_{c,a_i}(0)\;
\Big(\prod_{\dot\pi\ \rm legs}\frac{i\,k_i}{\rhohat}\Big)\,
\mathcal J^{\rm (v)}_{-\vec a}\Big]\;+\;2\ \text{perms}\,,
\qquad
\begin{array}{c|cccc}
v & \sigma^{3} & \dot\pi_c\sigma^{2} & \dot\pi_c^{2}\sigma & \dot\pi_c^{3}\\
\hline
n & 0 & 1 & 2 & 3\\
\mathcal C_v & 4g & 2\alpha & 2/\Lambda_2 & 4\lambda_2
\end{array}
\end{equation}
where the two extra terms are the cyclic permutations of $(k_1,k_2,k_3)$,
and the $i/\rhohat$
prefactors of \eqref{eq:smlegdot} have been pulled out of the legs. The vertex integral
generalizes \eqref{eq:master}: picking one monomial per leg and using
$\int_0^{\infty}\d t\;t^{N-1}e^{-Kt}=\Ga(N)/K^{N}$,
\begin{equation}
\label{eq:smJgen}
\mathcal J^{\rm (v)}_{\vec a}
=\sum_{\rm monomials}\Big(\prod_{i=1}^{3}\tilde c_i\,k_i^{\,m_i}\Big)\,
\Ga(N)\iiint_1^{\infty}\prod_{i=1}^{3}\d\lambda_i\,
\lambda_i^{\,j_i}\,B_{a_i}(\lambda_i)\;\frac{1}{K^{N}}\,,
\qquad N=m_1+m_2+m_3-3+n\,,
\end{equation}
which for $\sigma^{3}$ has a single term, $N=3$, and reproduces
\eqref{eq:assembly}.

\paragraph*{Soft-leg and hard-leg integrals.} We again order $k_3\le k_2\le k_1$ and
write $1/K^{N}=k_3^{-N}(\lambda_3+\lamstar)^{-N}$.
To perform the integral over $\lambda_3$ using \eqref{eq:smtowers}, we expand $\lambda_3^{\,j_3}=\big[(\lambda_3+\lamstar)-\lamstar\big]^{j_3}$ with the binomial theorem to get
\begin{equation}
\label{eq:smsoftred}
\int_1^{\infty}\!\d\lambda_3\;\lambda_3^{\,j_3}\,
\frac{B_{a_3}(\lambda_3)}{(\lambda_3+\lamstar)^{N}}
=\sum_{i_3=0}^{j_3}\binom{j_3}{i_3}(-\lamstar)^{\,j_3-i_3}\,
T_{N-i_3}(\lamstar)\,,
\end{equation}
with $T_\beta$ the soft-leg integral \eqref{eq:softblock} evaluated on channel $a_3$.
Once the series \eqref{eq:smtowers} of each $T_{N-i_3}$ is inserted, each term involves a hard-leg integral $\mathcal G_{a_1a_2}(p)$ \eqref{eq:smmaster}.
With the coefficients $t^{(\beta,a)}_{n}$ and
$t^{(\beta,a)}_{\pm,n}$ of \eqref{eq:smtcoefs}, the vertex integral reads
\begin{equation}
\label{eq:smJassembled}
\boxed{\;\begin{aligned}
\mathcal J^{\rm (v)}_{\vec a}
={}&\sum_{\rm monomials}\Big(\prod_{i=1}^{3}\tilde c_i\,k_i^{\,m_i}\Big)\,
\Ga(N)\,k_3^{-N}
\sum_{i_3=0}^{j_3}\binom{j_3}{i_3}(-1)^{\,j_3-i_3}\\[2pt]
&\times\Bigg[\,\sum_{n\ge0}t^{(N-i_3,\,a_3)}_{n}\;
k_3^{\,p_{n}}\;\mathcal G_{a_1a_2}(p_{n})
\;+\;\sum_{\pm}\sum_{n\ge0}t^{(N-i_3,\,a_3)}_{\pm,n}\;
k_3^{\,p_{\pm,n}}\;\mathcal G_{a_1a_2}(p_{\pm,n})\Bigg]
\end{aligned}\;}
\end{equation}
with, for each $i_3$, $p_{n}=(N-i_3)+n-(j_3-i_3)$ and
$p_{\pm,n}=(N-i_3)-u_{\pm,n}-(j_3-i_3)$, and
$\mathcal G$ evaluated with Eq.~\eqref{eq:smmaster}.
Equations~\eqref{eq:smprefactors}--\eqref{eq:smJassembled}, with the definitions \eqref{eq:smtcoefs}, the moments \eqref{eq:smMn}, the residues \eqref{eq:smres} and the hard-leg integrals
\eqref{eq:smmaster}, are the complete result. Every coefficient is a closed
expression in the parameters of the theory.

\end{document}